\documentclass[conference]{IEEEtran}
\IEEEoverridecommandlockouts
\usepackage{cite}
\usepackage{amsmath,amssymb,amsfonts}
\usepackage{algorithmic}
\usepackage{graphicx}
\usepackage{textcomp}
\usepackage{xcolor}

\newtheorem{theorem}{Theorem}

\newtheorem{proposition}{Proposition}

\newtheorem{conjecture}{Conjecture}

\def\BibTeX{{\rm B\kern-.05em{\sc i\kern-.025em b}\kern-.08em
    T\kern-.1667em\lower.7ex\hbox{E}\kern-.125emX}}
\begin{document}

\title{Existence of balanced functions that are not derivative of
bent functions
\thanks{The research  has been carried out within the framework of a state
assignment of the Ministry of Education and Science of the Russian
Federation for the Institute of Mathematics of the Siberian Branch
of the Russian Academy of Sciences (project no. FWNF-2022-0017.} }

\author{\IEEEauthorblockN{ Vladimir N. Potapov}
\IEEEauthorblockA{\textit{Sobolev Institute of Mathematics } \\
Novosibirsk, Russia \\
vpotapov@math.nsc.ru}  }

\maketitle

\begin{abstract}
It is disproved the conjecture (\cite{Tok16},\cite{Shap}) that any
balanced boolean function of appropriate degree is a derivative of
some bent function. This result is based on new upper bounds for the
numbers of bent and plateaued functions.
\end{abstract}


\section{Introduction}
Boolean bent  functions play a significant role in cryptography,
coding theory and combinatorics. They are the source of nonlinearity
and  provide some confusion in cryptosystems. Thus  investigations
of properties and diversity of bent functions are  important for
applications. Now the number of $n$-variable boolean bent functions
is known  for $n\leq 8$. For even $n\leq 10$ there are only lower
and upper bounds for the number of bent functions. Until recently,
developments in the bounds for the number of bent functions were
asymptotically negligible  (see \cite{CK}, \cite {Tok},
\cite{Carlet}, \cite{CarMes}, \cite{Mes0}). Now new lower \cite{PTT}
and new upper \cite{Pot21}, \cite{Ag23}, \cite{Pot23+} bounds for
the number of bent functions were obtained.  But  a drastic gap
between the asymptotic upper and lower bounds still remains.

On the other side the new bounds allow us to answer some known
questions about properties of bent functions. There is a following
hypothesis.

\begin{conjecture}[\cite{Tok16}, p. 1] Any balanced  boolean function $f$ in an even
number of variables $n$ of degree at most $n/2 -1$ such that $f (x)
= f (x \oplus y)$ for every vector $x$ and some nonzero vector $y$
is a derivative of a bent function.
\end{conjecture}

Recently, this conjecture has been proved for a very particular
case.

\begin{proposition}[\cite{Shap}, Theorem 4]  Let $f$ be a quadratic
balanced boolean function in $n\geq6$ variables such that $f (x) = f
(x \oplus y)$ for every vector $x$ and some nonzero vector $y$. Then
$f$ is a derivative of a bent function in $n$ variables.
\end{proposition}

In this paper we refute Conjecture 1 in Theorem 1. Our proof is
based on the asymptotic upper bound for the number of $1$-plateaued
functions that are obtained by the restriction of bent functions on
a hyperplane (see \cite{Pot23+}).  It is possible but more
complicated to disprove Conjecture 1 based on upper bounds from
\cite{Pot21}, \cite{Ag23}. Moreover, we suppose that an approach
developed in Theorem 2 will be useful for verification of similar
hypotheses.

\section{Preliminaries}

Let $\mathbb{F}= \{0, 1\}$. The set $\mathbb{F}^n$ is called a
boolean hypercube (or a boolean $n$-cube). $\mathbb{F}^n$ equipped
with  coordinate-wise modulo 2 addition $\oplus$ can be considered
as an $n$-dimensional vector space. Functions
 $\phi_x(y)=(-1)^{\langle x,y\rangle}$ are called characters. Here $\langle
x,y\rangle=x_1y_1\oplus\dots\oplus x_ny_n$ is the inner product. Let
$G$ be a function that  maps from the boolean hypercube to real
numbers. The Fourier transform of $G$ is defined by the formula
$\widehat{G}(y)=(G,\phi_y)$, i.e., $\widehat{G}(y)$ are the
coefficients of the expansion of $G$ in the basis of characters.

We can define the Walsh--Hadamard transform of a boolean function
$f:\mathbb{F}^n\rightarrow \mathbb{F}$  by the formula
$W_f(y)=\widehat{(-1)^f}(y)$, i.e., $$W_f(y)=\sum\limits_{x\in
\mathbb{F}^n}(-1)^{f(x)\oplus \langle x,y\rangle}.$$ A boolean
function $b$ is called a bent function if $W_b(y)=\pm 2^{n/2}$ for
all $y\in \mathbb{F}^n$. It is easy to see that $n$-variable bent
functions  exist only if $n$ is even. A boolean function $p$ is
called an $s$-plateaued function if $W_p(y)=\pm 2^{(n+s)/2}$  or
$W_p(y)=0$ for all $y\in \mathbb{F}^n$. So, bent functions are
$0$-plateaued functions. $1$-Plateaued functions are called
near-bent.

From Parseval's identity  $$\sum\limits_{y\in \mathbb{F}^n}
\widehat{H}^2(y)= 2^{n}\sum\limits_{x\in \mathbb{F}^n}H^2(x),$$
where $H:\mathbb{F}^n\rightarrow \mathbb{C}$, it follows
straightforwardly:
\begin{proposition}\label{bentpla00}
For every  $s$-plateaued function,  a proportion of nonzero values
of its Walsh--Hadamard transform is equal to $\frac{1}{2^s}$.
\end{proposition}

We will use the following property of bent and plateaued functions
(see \cite{Carlet}, \cite{CarMes}, \cite{Mes0}). Let
$f:{\mathbb{F}}^n\rightarrow {\mathbb{F}}$ be an $s$-plateaued
function, let $A:{\mathbb{F}}^n\rightarrow {\mathbb{F}}^n$ be a
non-degenerate affine transformation and let
$\ell:{\mathbb{F}}^n\rightarrow {\mathbb{F}}$ be an affine function.
Then $g=(f\circ A)\oplus\ell$ is an $s$-plateaued  function. The
functions $f$ and $g$  are called AE-equivalent.

It is well known (see e.g. \cite{Carlet},\cite{Tsf}) that for any
function $H,G:\mathbb{F}^n\rightarrow \mathbb{C}$ it holds
$$\widehat{H*G}= {\widehat{H}\cdot\widehat{ G}} \qquad {\rm
and}\qquad \widehat{(\widehat{H})}=2^nH,$$ where
$H*G(z)=\sum\limits_{x\in \mathbb{F}^n}H(x)G(z\oplus x)$ is a
convolution. Consequently, it holds
\begin{equation}\label{equpperbent11}
2^{n}H*G= \widehat{\widehat{H}\cdot\widehat{ G}} \ \mbox{and}\
\widehat{H}*\widehat{ G}= 2^{n}\widehat{H\cdot G}.
\end{equation}
 Let $\Gamma$ be a subspace of the hypercube.
  Denote by
   $\Gamma^\perp$  a
dual subspace, i.e., $\Gamma^\perp=\{y\in \mathbb{F}^n : \forall
x\in \Gamma, \langle x,y\rangle=0 \}.$ Let ${\bf1}_S$ be an
indicator function for $S\subset\mathbb{F}^n $. It is easy to see
that for every subspace $\Gamma$ it holds
$\widehat{{\bf1}_{\Gamma^\perp}}=2^{n-\mathrm{dim}\,\Gamma}{\bf1}_{\Gamma}$.
By (\ref{equpperbent11}) we have
\begin{equation}\label{equpperbent1}
{H}*{\bf1}_{\Gamma^\perp}=2^{-\mathrm{dim}\,\Gamma}\widehat{\widehat{H}\cdot\mathbf{1}_{\Gamma}}
\end{equation}
for any subspace $\Gamma\subset \mathbb{F}^n$, $\bar{0}\in \Gamma$.
When we substitute  $\Gamma^\perp=\{\bar{0},a\}$ for some $a\in
\mathbb{F}^n$ in (\ref{equpperbent1}) we obtain
\begin{equation}\label{equpperbent10}
H(x)+H(x\oplus a)
=2^{-n+1}\widehat{\widehat{H}\cdot\mathbf{1}_{\Gamma}}.
\end{equation}

The  derivative of a boolean function $f$ in the direction of a
vector $a\in \mathbb{F}^n$ is defined as $D_af(x)= f(x)\oplus
f(x\oplus a)$. It is easy to see that $D_af(x)=D_af(x\oplus a)$. So,
$D_af$ is determined by its restriction on the corresponding
hyperplane. If $e$ is a unit vector from the standard basis then we
will consider $D_ef$ as a boolean function in $n-1$ variables.
Without loss of generality (see EA-equivalence) we will consider
derivatives of bent functions only by coordinate directions.

\section{Main results}

Denote by $\mathrm{supp}(G)=\{x\in \mathbb{F}^n : G(x)\neq 0\}$ the
support of $G$.  We need the following known property of bent
functions (see e.g.\,\cite{Mes0}).

\begin{proposition}\label{bentpla1}
Let $f$ be an $n$-variable bent function and let $\Gamma$ be a
hyperplane. Consider $h={f}\cdot{\bf1}_{\Gamma}$ as an
$(n-1)$-variable function. Then $h$ is a  $1$-plateaued function.
\end{proposition}

\begin{proposition}\label{shap1}
Let $f$ be an $n$-variable bent function. Then the support of $D_ef$
is the complement of the support of the Walsh--Hadamard transform of
a $1$-plateaued function  obtained by the restriction of some
$n$-variables bent function on the hyperplane.
\end{proposition}
Proof. Let $H=(-1)^{f}$. It is easy to see that $D_ef(x)=0$ if and
only if $H(x)+H(x\oplus e)\neq 0$. By  (\ref{equpperbent10}) it
holds $\mathrm{supp}(D_ef)=
\mathbb{F}^{n-1}\setminus\mathrm{supp}(\widehat{\widehat{H}\cdot\mathbf{1}_{\Gamma}})$.
It is well known that $\widehat{H}=(-1)^g$ for a bent function $g$
dual to $f$. By Proposition \ref{bentpla1},
$g\cdot\mathbf{1}_{\Gamma}$ is a $1$-plateaued  function. Then
$\widehat{\widehat{H}\cdot\mathbf{1}_{\Gamma}}=\widehat{(-1)^g\cdot\mathbf{1}_{\Gamma}}$
is the Walsh--Hadamard transform of the  $1$-plateaued function.
$\square$

Every boolean function $f$ can be represented as a boolean
polynomial by the only way. It holds
\begin{equation}\label{equpperbent00}
 f(x_1,\dots,x_n)=\bigoplus\limits_{y\in
    \mathbb{F}^n}g(y)x_1^{y_1}\cdots x_n^{y_n},
\end{equation}
     where $x^0=1, x^1=x$, $g:\mathbb{F}^{n}\rightarrow \mathbb{F}$
     is a the M\"obius transform of $f$.
The maximal degree of this polynomial   is called the algebraic
degree of $f$. A boolean function is called balanced if it takes
values $0$ and $1$ equal times. By Proposition \ref{bentpla00} the
support of the the Walsh--Hadamard transform of every $1$-plateaued
function is a balanced function.

Let $\mathcal{N}_0(n,1)$ be the binary logarithm of the number of
$n$-variable $1$-plateaued  functions  which are obtained by the
restriction of  $(n+1)$-variable bent functions on a hyperplane.
Note that the $1$-plateaued function from Proposition \ref{shap1} is
such a restriction. We use the following upper bound for the number
of $1$-plateaued  functions.

\begin{proposition}[\cite{Pot23+},Corollary 4]\label{number_plat}\\
$\mathcal{N}_0(n-1,1)< 3.47\cdot2^{n-4}(1+o(1))$ as
$n\rightarrow\infty$.
\end{proposition}

\begin{theorem}
For sufficiently large even $n$, there exists an $(n-1)$-variable
balanced boolean function of degree not greater than $n/2-1$ such
that is not equal to a derivative of any bent function.
\end{theorem}
Proof. Firstly, we will find a lower bound for the number of
$n$-variable balanced boolean functions with degree $n/2$. Denote by
${\mathrm wt}(f)$ the weight of boolean function $f$, i.e.,
${\mathrm wt}(f)=|\{ x\in \mathbb{F}^n  \ |\ f(x)=1 \}|$. Let
$A^k_n(t)$ be the set of  $n$-variable boolean functions with degree
not greater than $k$ and weight $t$. It is easy to see that
$|A^k_n(t)|=|A^k_n(2^n-t)|$ and by (\ref{equpperbent00}) it holds
\begin{equation}\label{red23n1}
\sum\limits_{t=0}^{2^{n}}|A^{k}_{n}(t)|=2^{\sum\limits_{m=0}^{k}{n
\choose m}}.
\end{equation}
  If $f\in A^k_n(t_1)$ and $g\in A^k_n(t_2)$ then
$$f(x_1,\dots,x_{n})x_{n+1}\oplus g(x_1,\dots,x_{n})(x_{n+1}\oplus
1)$$ $$ =h(x_1,\dots,x_{n},x_{n+1})\in A^{k+1}_{n+1}(t_1+t_2).$$
Moreover,  such decomposition is true for any  $h\in
A^{k+1}_{n+1}(t)$. Therefore,

$$|A^k_n(2^{n-1})|=
\sum\limits_{t=0}^{2^{n-1}}|A^{k-1}_{n-1}(t)||A^{k-1}_{n-1}(2^{n-1}-t)|$$
\begin{equation}\label{red23n2}
= 2\sum\limits_{t=0}^{2^{n-2}}|A^{k-1}_{n-1}(t)|^2.
\end{equation}

By the Cauchy-Schwarz inequality and (\ref{red23n2}) we have
$$|A^k_n(2^{n-1})|\geq
2\left(\sum\limits_{t=0}^{2^{n-2}}|A^{k-1}_{n-1}(t)|\right)^2/(2^{n-2}+1)^2.$$
By (\ref{red23n1}) we obtain $$|A^{n/2}_n(2^{n-1})|\geq
\frac{2(2^{2^{n-2}}\cdot \frac12)^2}{2^{2n-4}+2^{n-1}+1}\geq
2^{2^{n-1}-1}/2^{2n-2}.$$ Thus, the number of $(n-1)$-variable
balanced boolean functions of degree $n/2-1$ is not less than
$$ 2^{2^{n-2}(1+o(1))}>
2^{3.47\cdot2^{n-4}}. $$ Consequently, for sufficiently large even
$n$ the number of $1$-plateaued functions  obtained by the
restrictions of  $n$-variables bent functions on hyperplanes is less
than the number of balanced functions of degree $n/2-1$ by
Proposition \ref{number_plat}. By Proposition \ref{bentpla1} it is
completed the proof. $\square$

In the development of this topic it is natural to study the
following question. What boolean functions could be a support of the
Walsh--Hadamard transform of some boolean function?

\begin{theorem}
For any $n\geq k>0$ there exists a balanced $n$-variable boolean
function of degree not greater than $k$ such that is not equal to
the support of the Walsh--Hadamard transform of some $n$-variable
boolean function of degree not greater than  $k$.
\end{theorem}
Proof. Let $E_k$ be the number of balanced $n$-variable boolean
function of degree $k$. The total number of zero-values of such
functions is equal to $E_k2^{n-1}$.  Since the degree of
$f(x)\oplus\langle x,a\rangle$ is equal to the degree of $f$, it
holds
$$E_k=|\{f:\mathbb{F}^n \rightarrow \mathbb{F} \ |\
\widehat{(-1)^f}(0)=0, {\rm deg}(f)\leq k\}|$$ $$= |\{f:\mathbb{F}^n
\rightarrow \mathbb{F} \ |\ \widehat{(-1)^f}(a)=0, {\rm deg}(f)\leq
k\}|.$$ By double counting we obtain that the total number of zeros
in the Walsh--Hadamard transform of all $n$-variable boolean
functions of degree not greater than  $k$ is equal to $E_k2^{n}$.

The supports of the Walsh--Hadamard transform of $g$ and $g\oplus 1$
coincide. Then if a balanced boolean function $f$ is the support of
the Walsh--Hadamard transform of some function $g$ then it is true
for
  at least two
functions $g$ and $g\oplus 1$. Therefore all zeros of the
Walsh--Hadamard transforms could be utilized by balanced functions.
But there are a lot of Walsh--Hadamard transforms with the other
numbers of zeros. For example, the Walsh--Hadamard transform of
constants.
 $\square$

\end{document}